\documentclass[]{emulateapj-rtx4}

\usepackage{natbib,color} 
\bibpunct{(}{)}{;}{a}{}{,}

\DeclareRobustCommand{\ion}[2]{%
\relax\ifmmode
\ifx\testbx\f@series
{\mathbf{#1\,\mathsc{#2}}}\else
{\mathrm{#1\,\mathsc{#2}}}\fi
\else\textup{#1\,{\mdseries\textsc{#2}}}%
\fi}

\shorttitle{Polarization signature of magnetic fields in simulations and observations}
\shortauthors{Beck, C.; Fabbian, D.; Rezaei, R.; Puschmann, K.G.}

\begin{document}
\title{The polarization signature of photospheric magnetic fields in 3D MHD simulations and observations at disk center} 


\author{C. Beck}
\affil{National Solar Observatory}
\author{D. Fabbian}
\affil{Max-Planck-Institut f\"ur Sonnensytemforschung}
\author{R. Rezaei}
\affil{Instituto de Astrof\'{\i}sica de Canarias}
\affil{Departamento de Astrof{\'i}sica, Universidad de La Laguna}
\author{K.G. Puschmann}
\affil{Germany}



\begin{abstract}
Before using three-dimensional (3D) magneto-hydrodynamical (MHD) simulations of the solar photosphere in the determination of elemental abundances, one has to ensure that the correct amount of magnetic flux is present in the simulations. The presence of magnetic flux modifies the thermal structure of the solar photosphere, which affects abundance determinations and the solar spectral irradiance. The amount of magnetic flux in the solar photosphere also constrains any possible heating in the outer solar atmosphere through magnetic reconnection. We compare the polarization signals in disk-center observations of the solar photosphere in quiet-Sun regions with those in Stokes spectra computed on the basis of 3D MHD simulations having average magnetic flux densities of about 20, 56, 112 and 224\,G. This approach allows us to find the simulation run that best matches the observations. The observations were taken with the \textit{Hinode SpectroPolarimeter} (SP), the \textit{Tenerife Infrared Polarimeter} (TIP), the \textit{Polarimetric Littrow Spectrograph} (POLIS) and the \textit{GREGOR Fabry-P{\`e}rot Interferometer} (GFPI), respectively. We determine characteristic quantities of full Stokes profiles in a few photospheric spectral lines in the visible (630\,nm) and near-infrared (1083 and 1565\,nm). We find that the appearance of abnormal granulation in intensity maps of degraded simulations can be traced back to an initially regular granulation pattern with numerous bright points in the intergranular lanes before the spatial degradation. The linear polarization signals in the simulations are almost exclusively related to canopies of strong magnetic flux concentrations and not to transient events of magnetic flux emergence. We find that the average vertical magnetic flux density in the simulation should be less than 50\,G to reproduce the observed polarization signals in the quiet Sun internetwork. A value of about 35\,G gives the best match across the SP, TIP, POLIS and GFPI observations. 
\end{abstract}


\section{Introduction}
For the derivation of elemental abundances in the solar photosphere, numerical
hydrodynamical (HD) simulations have been routinely used during the last decade \citep[\textit{e.g.},][and references therein]{allendeprieto+etal2001a,asplund+etal2005a}. The thermal structure in such HD simulations is determined by the convective energy transport in the uppermost layers of the solar convection zone, near the transition to an optically thin atmosphere with radiative energy transport. The presence of magnetic flux modifies the thermal structure and the behavior of the granulation pattern, and hence can impact the determination of abundances from spectra synthesized from simulations \citep[][]{fabbian+etal2010,fabbian+etal2012,fabbian+etal2015}. Before using more realistic magneto-hydrodynamical (MHD) simulations in abundance determinations, one has first to ensure that the chosen MHD simulations reproduce the observed properties of magnetic fields in the solar photosphere. 

The amount of magnetic flux in the solar photosphere in the so-called quiet Sun (QS) outside of active regions has also implications for other topics. On the one hand, it constrains any possible heating of the outer solar atmosphere by magnetic reconnection \citep[\textit{e.g.},][]{wiegelmann+etal2014,guerreiro+etal2015,chitta+etal2017}. During the minima of solar activity, most to all of the solar disk corresponds to QS. Any heating of the solar chromosphere, transition region or corona by magnetic reconnection during an activity minimum cannot exceed the magnetic flux energy present at that stage in the photosphere. On the other hand, the presence and amount of magnetic flux in the QS photosphere also affects the solar spectral irradiance \citep{criscuoli+uitenbroek2014,faurobert+etal2016}. An accurate value, or at least an upper limit on the characteristic amount of magnetic flux in QS regions is thus helpful to estimate the maximal cycle variability at different atmospheric layers of the Sun and to constrain the basal activity level during solar minima.

Comparisons between observations and MHD simulations have been used in the past to determine the average magnetic flux density in the magnetically least active regions of the solar surface, the internetwork (IN) QS \citep[\textit{e.g.},][]{khomenko+etal2005,martinezgonzalez+etal2008,danilovic+etal2010}. The direct derivation of the average magnetic flux density from spectropolarimetric observations using the Zeeman effect is hindered by ambiguities between magnetic field properties and their signature in observed polarization profiles. On the one hand, spectral lines in the visible wavelength range form in the so-called weak-field limit (WFL), in which the wavelength splitting of the Zeeman components is still smaller than the Doppler and thermal broadening of the lines \citep[\textit{e.g.},][]{jefferies+etal1989}. On the other hand, the magnetic fields are usually confined to narrow intergranular lanes, because in the solar photosphere the energy of the dynamical mass motions dominates over magnetic forces. Photospheric magnetic fields are therefore commonly not fully spatially resolved \citep[][and references therein]{navarro+almeida2003,lites+socasnavarro2004,dewijn+etal2009} in current observations with a typical spatial resolution between 0\farcs1 \citep{scharmeretal08} and 1$^{\prime\prime}$ \citep{beck+rezaei2009}. 

In addition, the observed polarization signal also depends on the thermodynamic properties of the atmosphere because the polarization signal and its measurement are always only relative to the total intensity. This all together yields an ambiguity between the magnetic field strength, magnetic flux density, the magnetic filling factor inside a spatial pixel and the thermodynamic state of the atmosphere,  \textit{i.e.}, the temperature stratifications both inside and outside of the magnetized plasma in the case of unresolved fields. In consequence, divergent results on the average vertical magnetic flux density in the IN have been obtained depending on which spectral line and which method was used in the derivation \citep[\textit{e.g.},][]{grossmanndoerth+etal1996,cerdena+etal2003,navarro+lites2004,khomenko+etal2005,orozco+etal2007a,lites+etal2008,martinezgonzalez+etal2008,beck+rezaei2009,stenflo2010,danilovic+etal2010a,danilovic+etal2016}. The numbers range from less than 10\,G to about 50\,G, with a generic value of about 20\,G for the vertical magnetic flux density on average \citep[see][their Figure~3]{sanchezalmeida+martinezgonzalez2011}. Determinations of the average magnetic flux density in the solar photosphere via the Hanle effect that is (partly) insensitive to the problem of the spatial resolution yielded a significantly larger value of up to 100\,G \citep{trujillobueno+etal2004}, as also found in \citet{lites+etal2008} for the transversal magnetic flux density.

Direct comparisons between synthetic spectra from numerical MHD simulations
and observed spectra as for instance done in \citet{khomenko+etal2005} provide an elegant solution to avoid the ambiguities in the derivation of magnetic field properties from observations. One drawback of this approach is, however, that one has to rely on statistical quantities because one never has a direct relation between individual spectra in simulations and observations. \citet{carroll+kopf2008} presented one of the few cases where MHD simulations were used as a basis of a direct analysis of observed spectra in an inversion approach using neural networks. The approach of using MHD simulations in the inversion was taken up recently by \citet{riethmueller+etal2016}. 

In the current contribution, we investigate the polarization signature of photospheric magnetic fields in MHD simulations and observations taken at the center of the solar disk to determine which simulation best matches the observations in different spectral lines. The observations, the MHD simulation runs, and the degradation of the simulations in spatial and spectral resolution are described in Section \ref{sect_obs}. Section \ref{sect_param} gives the characteristic quantities of the polarization signal that we employ for the comparison. Our results are given in Section \ref{sect_results} and are discussed in Section \ref{sect_disc}. Section \ref{sect_concl} provides our conclusions.

\section{Observations, MHD runs, and Degradation of Simulations}\label{sect_obs}
The complete set of observations is described in detail in \citet[][BE13
in the following]{beck+etal2013}. Each data set consists of 
spectropolarimetric observations of QS at or near the center of the solar
disk. The data were taken with the \textit{SpectroPolarimeter} (SP) onboard of the \textit{Hinode} satellite \citep{kosugi+etal2007,tsuneta+etal2008}, the \textit{GREGOR Fabry-P{\`e}rot Interferometer} \citep[GFPI;][see also \citeauthor{puschmann2016} \citeyear{puschmann2016} and references therein]{puschmann+etal2006}, the \textit{POlarimetric LIttrow Spectrograph} \citep[POLIS;][]{beck+etal2005b}, and the \textit{Tenerife Infrared Polarimeter} \citep[TIP;][]{collados+etal2007}. The spectral lines observed with the latter were \ion{Si}{i} at 1082.7\,nm, \ion{Fe}{i} at 1564.8\,nm and \ion{Fe}{i} at 1565.2\,nm, whereas all other data covered one or both of the \ion{Fe}{i} lines at 630.15\,nm and 630.25\,nm. 
\begin{table}
\caption{Rms noise ($\equiv 1 \sigma$) in $I/I_c$, $Q/I_c$, $V/I_c$ and $p/I_c$ in units of $10^{-3}$.\label{tab1}}
\begin{tabular}{c|cccccc}
instrument & SP & GFPI  & POLIS & TIP & TIP \cr\hline\hline
$\lambda$  & 630\,nm & 630.25\,nm  &  630\,nm & 1083\,nm &
1565\,nm\cr
$I$  & 10 &  12  & 6.7 & 2.9 & 2.0 \cr
$Q$   & 2.6 & 1.3  & 1.3 & 1.4 & 0.5\cr
$V$   & 2.4 & 1.6  & 1.0 & 2.0 & 0.5 \cr
$p$  & 1.8 & 1.3  & 0.8 & 1.3 & 0.3 \cr
\end{tabular}
\end{table}

\begin{figure*}
\begin{minipage}{12cm}
\resizebox{11cm}{!}{\includegraphics{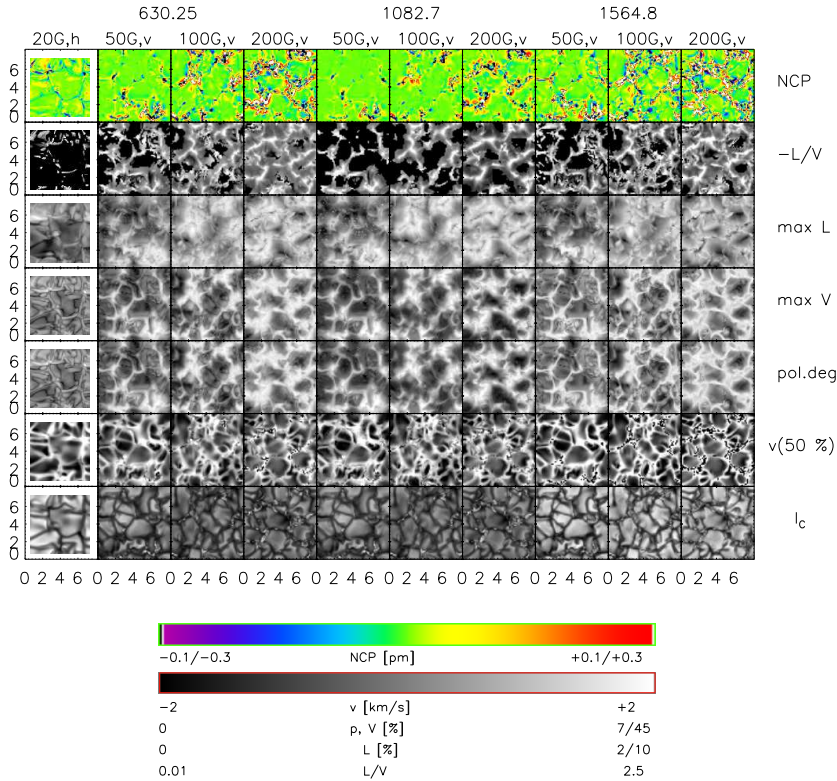}}
\end{minipage}
\begin{minipage}{5.6cm}
\caption{Maps of polarization and intensity parameters for the simulation
  runs at full resolution. Bottom to top: continuum intensity, bisector
  velocity at 50\,\% line depth, polarization degree, maximum Stokes $V$
  amplitude, maximum linear polarization $L$, ratio $L/V$ and net circular polarization. The values of $p$, $V$ and $L$ are displayed on a logarithmic scale, while $L/V$ is shown in a linear inverted color scale. Regions with $V$ signals below 0.1\,\% were suppressed in the latter (uniform black areas). {Four left-most columns}: results based on spectra at 630.25\,nm for the 20 G h, 50 G v, 100 G v, and 200 G v runs. Fifth to seventh column: results based on spectra at 1082.7\,nm for the 50 G v, 100 G v, and 200 G v runs. Eighth to tenth column: results based on spectra at 1564.8\,nm for the 50 G v, 100 G v, 200 G v runs. The continuum intensity is displayed within its variable full range each time. The display ranges for the 20 G h run are given by the numbers of smaller modulus for $p$, $V$, $L$, and the NCP in the corresponding labels of the legend color bars. Tick marks are in arcsec.} \label{mhd2d_fullres}
\end{minipage}
\end{figure*}

In difference to BE13 we replaced the TIP data set at 1565\,nm with the second observation of \citet{beck+rezaei2009}. This observation covered the same field of view (FOV) as the one in BE13, but had an integration time of about 30\,s at otherwise identical settings. We also use here only the GFPI data that were reduced by means of Multi-Object Multi-Frame Blind Deconvolution \citep[MOMFBD;][]{vannoort+etal2005}, a further option for image reconstruction and spectral line deconvolution implemented in the GFPI \textit{imaging Parallel Organized Reconstruction Data Pipeline} \citep[iSPOR-DP;][]{puschmann+beck2011,puschmann2016}. The spatial and spectral sampling of the individual observations are listed in Table 1 of BE13, we only add here in Table \ref{tab1} the root-mean-square (rms) values of the noise in continuum windows of the polarization profiles. 

For the comparison of observed and synthetic polarization profiles, we used synthetic spectra computed on the basis of four different MHD simulation runs. Three of the MHD runs were performed with the \textit{Stagger} code \citep[see, \textit{e.g.},][and references therein]{fabbian+etal2012,beeck+etal2012} and are described in more detail in \citet[][]{fabbian+etal2010}. They differ in the amount of magnetic flux introduced into a thermodynamically relaxed HD simulation run, namely, an initial average unipolar vertical magnetic flux density of about 56, 112, and 224\,G was used. In the following, we label these runs as ``\textit{50 G v}'', ``\textit{100 G v}'', and ``\textit{200 G v}'' for simplicity. The fourth MHD run was performed with the \textit{COnservative COde for the COmputation of COmpressible COnvection in a BOx of L Dimensions} \citep[CO$^5$BOLD;][]{freytag+etal2012}, with an initial horizontal magnetic flux density of 20\,G and is described in \citet[][{see also \citeauthor{steiner+etal2008} \citeyear{steiner+etal2008}}]{schaffenberger+etal2005,schaffenberger+etal2006}. For this run, labeled as ``\textit{20 G h}'', we have no synthetic spectra of the \ion{Si}{i} line at 1082.7\,nm available. From each of the four MHD runs, we selected one temporal snapshot that belonged to the statistically-stationary regime as input for the spectral synthesis. A version with four times fewer grid points in each horizontal direction was used for the \textit{Stagger} simulations to reduce the computational effort (see Section 3 of BE13 for more details).

We then applied to the resulting MHD synthetic spectra the same spatial and spectral degradation as in BE13, where the appropriate values were determined by matching results based on a \textit{Stagger} code HD simulation to observed spectra. The differences between the HD and MHD simulation runs in terms of, \textit{e.g.}, rms intensity contrasts are rather small (see Table 4 of BE13), hence the degradation values derived from the HD case should also apply for the  MHD runs. For the 20\,G h run, we repeated the determination of the spatial degradation as described in BE13 because of the different spatial sampling (0\farcs055 per pixel) in this simulation. In the following, ``\textit{degraded simulations}'' denotes spatially and spectrally degraded simulations with additionally added noise corresponding to the noise level in the observations (Table \ref{tab1}), whereas ``\textit{simulations at full resolution}'' refers to the original simulations' spectra without any degradation or addition of noise. 

\begin{figure*}
\resizebox{17.6cm}{!}{\hspace*{.5cm}\includegraphics{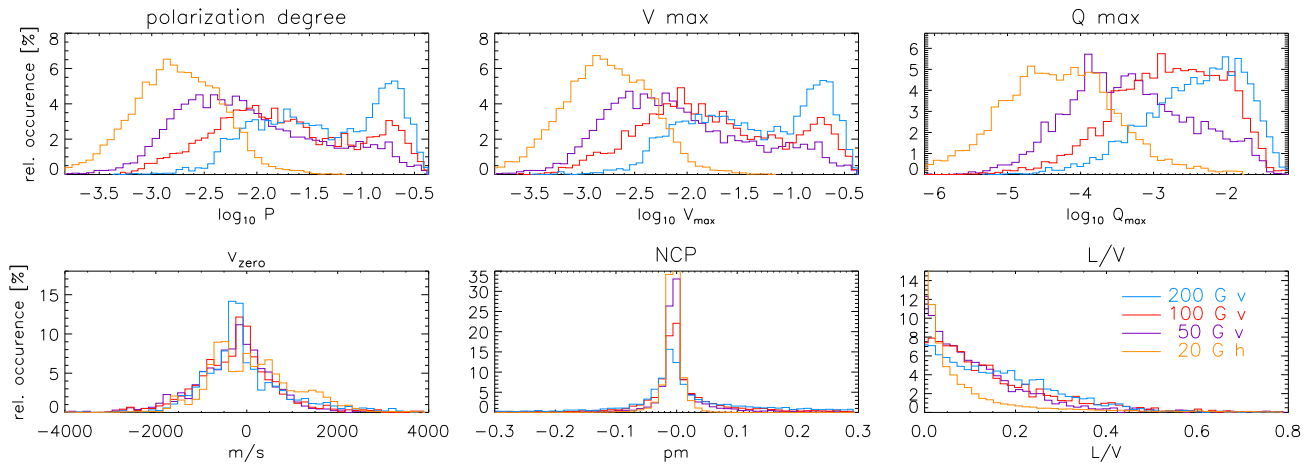}}
\caption{Statistics of polarization signals of the 630.25\,nm line in the simulations at full resolution. Top row, left to right: histograms of polarization degree, maximum Stokes $V$ amplitude, and maximum Stokes $Q$ amplitude. Bottom row, left to right: histograms of zero-crossing velocity, NCP, and $L/V$ ratio. Orange/purple/red/blue lines: 20 G h, 50 G v, 100 G v, 200 G v runs.\label{stat1_fullres}}
\end{figure*}
\section{Characteristic Polarization Parameters}\label{sect_param}
In addition to the line parameters derived from the intensity profile Stokes
$I$ in BE13, we determined the following list of line parameters from the
polarization profiles Stokes $QUV$ \citep[see][]{sigwarth+etal1999,beck+etal2007}:
\begin{itemize}
\item The maximum Stokes $QUV/I_c$ amplitudes.
\item The maximum of the total linear polarization $L = \sqrt{Q^2+U^2}$ and of the polarization degree $p=\sqrt{Q^2+U^2+V^2}$. 
\item The ratio $L/V$ around the wavelength of maximum $V$ signal as an indicator of the inclination of the magnetic field to the line of sight (LOS).
\item The net circular polarization (NCP) from an integration of the Stokes $V$
  signal. We used NCP\,=\,${\rm sign} \cdot \int V(\lambda) d\lambda$, where the sign ($\pm 1$) of the $V$ signal was defined such as to yield the area of the blue $V$ lobe minus that of the red one. 
\item The zero-crossing velocity of the Stokes $V$ signal as a measure of the
  Doppler shift in the magnetized plasma.
\end{itemize}

These parameters were derived separately for every spectral line in each individual set of Stokes profiles for both observations and simulations.

\section{Results}\label{sect_results}
\subsection{MHD Simulations at Full Resolution}
Figure \ref{mhd2d_fullres} shows overview maps of various
thermodynamic and polarimetric line parameters for the MHD simulation runs at
full resolution. The synthetic spectra of three lines (630.25\,nm, 1082.7\,nm,
1564.8\,nm) were employed. The presence of the magnetic fields leaves its
signature both in the continuum intensity (bottom row) and the bisector
velocity at 50\,\% line depth (second row). In the former, the number
density of magnetic bright points \citep[BPs; \textit{e.g.},][]{berger+title2001,almeida+etal2004,shelyag+etal2004,ishikawa+etal2007,beck+etal2007,sanchezalmeida+etal2010,viticchie+etal2010} in the intergranular lanes increases with increasing magnetic flux, whereas in the 20 G h run they are absent. The size of granules reduces with increasing magnetic flux in the maps of the LOS velocity. Several small-scale upflow features that correspond to fragmented granules appear for the 100 G v and 200 G v runs. 

A comparison of the polarization degree (third row from the bottom) and the maximum Stokes $V$ amplitude (fourth row) reveals that the polarization degree is dominated by the LOS magnetic flux, \textit{i.e.}, vertical magnetic fields. In several cases, strong polarization signals completely surround individual granules for all of the \textit{Stagger} MHD runs employed here.

The comparison of the fourth to sixth rows that display the circular and linear polarization signal and the ratio $L/V$, respectively, shows that the linear polarization signals are mainly related to the canopy of laterally expanding concentrations of magnetic flux. This is clearly visible  in case of the 100 G v and 200 G v runs, where very low values of $L/V$ (white areas in the inverted grey scale) correspond to intergranular lanes (IGLs) with strong total polarization signal. The locations with significant linear polarization signal then form a halo of decreasing amplitude around the IGLs with the latter showing at their very center a dark lane of reduced linear polarization signal. The NCP (top row) also confirms the existence of magnetic canopies, showing negative values above the central parts of the flux concentrations and halos of positive NCP surrounding them. This behavior of the NCP was identified as typical for the lateral expansion of magnetic flux concentrations embedded in a convective surrounding \citep[\textit{e.g.},][]{rezaei+etal2007,beck+etal2010,martinezgonzalez+etal2012}. Significant NCP values are limited to the close surroundings (about 1$^{\prime\prime}$) of strong polarization signals.

Figure \ref{stat1_fullres} shows histograms of various polarization parameters
of the \ion{Fe}{i} line at 630.25\,nm for the simulations at full
resolution. The histograms of the polarization degree, Stokes $V$ and Stokes $Q$ amplitudes (top row) again confirm the dominant role of the circular polarization signals for the total polarization degree, whereas the linear polarization signals are usually an order of magnitude weaker. The increase of the magnetic flux density shifts the histograms successively to larger polarization amplitudes, with a clear bump at polarization degrees or Stokes $V$ amplitudes of about 15\,\% (log$_{10} \sim -0.75$) for the 100 G v and 200 G v runs. These strong polarization signals correspond to flux concentrations inside IGLs which can only form if sufficient magnetic flux is present in the simulation box. The zero-crossing velocities (lower left panel) span a range of about $\pm 2$\,kms$^{-1}$, with an average value close to zero (between about $\pm 100$\,ms$^{-1}$). Both the zero-crossing velocity and the NCP histograms show only minor variations with the magnetic flux density in the simulation box, whereas the histogram of the latter exhibits a slight broadening with increasing magnetic flux density. A similar weak trend of having broader distributions with increasing magnetic flux density is seen in the $L/V$ ratio (rightmost bottom panel). 

\begin{figure}
\resizebox{8.8cm}{!}{\includegraphics{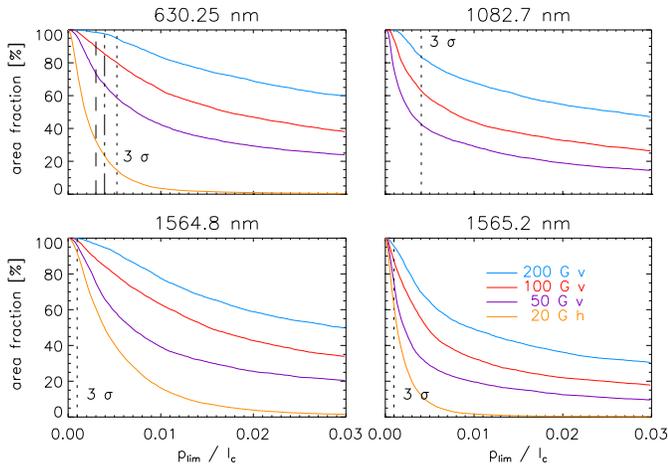}}
\caption{Area fraction of profiles with a polarization degree above the threshold $p_{\rm lim}$ for the simulations at full resolution. {Clockwise, starting
    left top}: 630.25\,nm, 1082.7\,nm, 1565.2\,nm, and
  1564.8\,nm spectra. The vertical lines indicate the 3-$\sigma$ significance
  level of the corresponding observations. For the 630.25\,nm line, the dashed, triple-dot-dashed and dotted lines denote 3-$\sigma$ in POLIS, GFPI, and SP data, respectively. }\label{stat2_fullres}
\end{figure}

The area fraction of profiles above a given polarization threshold for the
simulations at full resolution is displayed in Figure \ref{stat2_fullres} for
all spectral lines. This quantity corresponds to an integration of the histogram from the maximum value of polarization degree down to the threshold value. The (cumulative) area fraction is insensitive to the shape of the distribution, but measures the integrated area of the distribution. In this quantity, the four different simulation runs can clearly be distinguished by their respective curves. With increasing magnetic flux density, more and more profiles with large polarization degrees above 1\,\% appear. The curves of the area fraction in Figure \ref{stat2_fullres} gradually change from being nearly linear for the 200 G v run to a steeply-dropping and sharply-bending curve for the 20 G h run. The comparison of the curves with the respective 3-$\sigma$ significance levels of the observations (vertical lines) reveals that even without any degradation of the simulations, only the long-integrated high signal-to-noise (S/N) TIP data at 1565\,nm are able to detect the majority of the polarization signals. Depending on the instrument and the amount of magnetic flux in the simulations, 10\,\% to up to 80\,\% of the polarization signals in the simulations at full resolution cannot be detected in the observations \citep[cf.][]{beck+rezaei2009,puschmann+beck2011}. 
\begin{figure}
\resizebox{8.8cm}{!}{\includegraphics{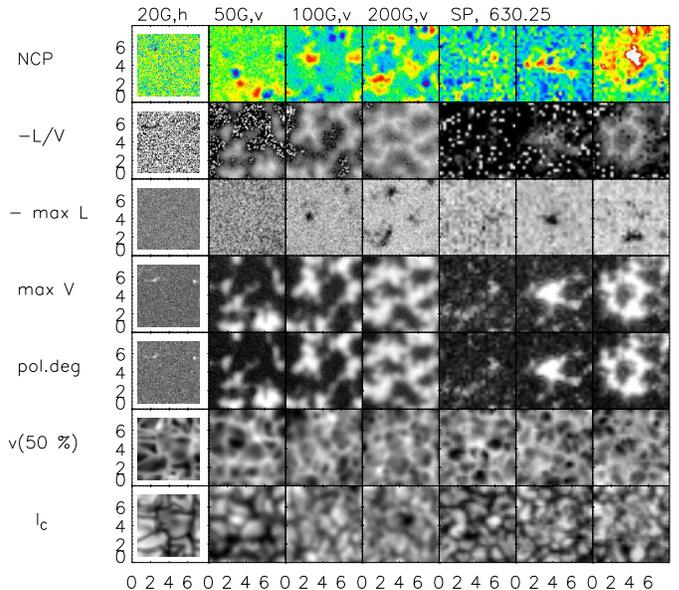}}
\caption{Maps of polarization and intensity parameters for the 630.25\,nm
  line in the SP observations and the simulations degraded to SP resolution. Same vertical layout as in Figure \ref{mhd2d_fullres}. The values of $p$ and $V$ are displayed on a logarithmic scale, while $L$ and $L/V$ are shown in a linear inverted color scale. Leftmost four columns: results based on degraded spectra of the 20 G h, 50 G v, 100 G v, and 200 G v runs. Fifth to seventh column: subfields in the SP data corresponding to IN, network, and plage-like magnetic fields. The locations of the subfields are marked in Figure \ref{sp_fullfov}. Tick marks are in arcsec.} \label{hinode2d}
\end{figure}
\begin{figure*}
\centerline{\resizebox{11.8cm}{!}{\includegraphics{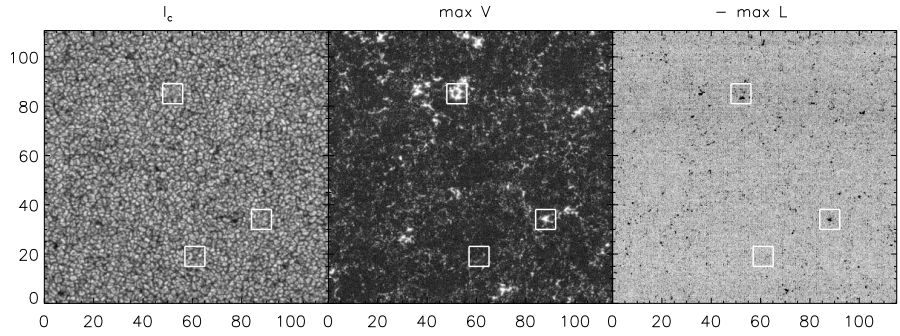}}}
\caption{Maps of continuum intensity (left panel), maximal $V$ amplitude (middle panel), and maximal linear polarization amplitude $L$ in inverted grey scale (right panel) for the 630.25\,nm line in the full FOV of the SP observations. The white squares indicate the subfields shown in Figure \ref{hinode2d}. Tick marks are in arcsec.}\label{sp_fullfov}
\end{figure*}
\subsection{SP Observations and Degraded MHD Simulations at 630\,nm}
Figure \ref{hinode2d} shows maps of the simulations degraded to the resolution of the SP data, together with three subfields of the full SP FOV that correspond to quiet IN (at $x,y \sim 60^{\prime\prime}, 20^{\prime\prime}$), network (at $x,y \sim 90^{\prime\prime}, 35^{\prime\prime}$), and a plage-like strong network area (at $x,y \sim 50^{\prime\prime}, 90^{\prime\prime}$). The locations of the subfields inside the SP FOV are indicated in Figure \ref{sp_fullfov} by white squares. Note that there are only very few regions in the full SP FOV with strong network or plage areas, thus the two respective subfields are not representative of the full FOV or internetwork QS in general.

\begin{figure*}
\resizebox{17.6cm}{!}{\includegraphics{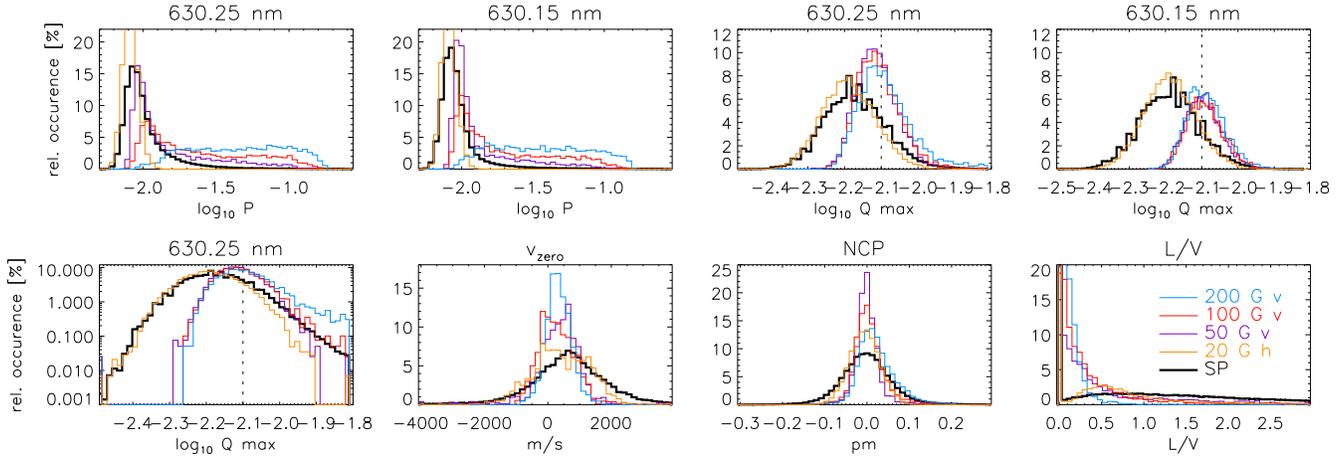}}
\caption{Statistics of polarization signals of the 630\,nm lines in the SP observations and the simulations degraded to SP resolution. Top row, left to right: histograms of the maximum polarization degree of 630.25\,nm and 630.15\,nm, and maximum Stokes $Q$ amplitude of 630.25\,nm and 630.15\,nm. Bottom row, left to right: histograms of the maximum Stokes $Q$ amplitude of 630.25\,nm with logarithmic $y$-axis, zero-crossing velocity, NCP, and $L/V$ ratio of 630.25\,nm. Black lines: observations. Orange/purple/red/blue lines: results based on degraded spectra of the 20 G h, 50 G v, 100 G v, and 200 G v runs.\label{sp_polstat}}\label{sp_ppolstat}
\end{figure*}

The increase of the magnetic flux density in the simulations and also in the three subfields of the observations clearly affects the continuum intensity
(bottom row of Figure \ref{hinode2d}). For the degraded intensity image of the 200 G v run or for the plage-like area in the observations, the granulation pattern partly disappears through the fading of the IGLs (``abnormal granulation'').  A comparison with the corresponding intensity image at full resolution in Figure \ref{mhd2d_fullres} reveals, however, that this effect is mainly related to the spatial degradation. The 200 G v simulation run at full resolution still shows a clear granulation pattern at 630.25\,nm, but with numerous BPs in the IGLs. The spatial degradation reduces the contrast of the IGLs because of the spatial smearing of the local intensity enhancements of the BPs, which creates the impression of abnormal granulation. 

As for the simulation runs at full resolution, the polarization degree in the
observations is dominated by the circular polarization signal. There are only a few isolated patches of significant linear polarization signal (fifth row from the bottom in Figure \ref{hinode2d}). The spatial distribution and area fraction of the linear polarization patches in observations and simulations are similar, \textit{i.e.}, no signal in the quiet IN and very few isolated patches in the subfields or simulations with large magnetic flux. We note that the linear polarization signal in all of the cases shown in Figure \ref{hinode2d} is directly related to vertical magnetic fields. Both the simulations at
full resolution (Figure \ref{mhd2d_fullres}) and at the spatial
resolution of the GFPI data (Figure \ref{gfpi_2d}) show that the linear
polarization patches in the simulations actually always trace the canopy of laterally expanding magnetic flux concentrations, with a reduction
of linear polarization signal in the center above the corresponding IGL
(especially clear in Figure \ref{gfpi_2d}). The map of linear polarization
signal across the full SP FOV (rightmost panel of Figure \ref{sp_fullfov}) exhibits a few additional cases of isolated linear polarization signal without any co-spatial or close-by circular polarization signal, but the majority of the linear polarization signal seems to be related to canopy magnetic fields \citep[cf.][]{ishikawa+tsuneta2011}. This is also confirmed by the maps of the  $L/V$ ratio in the sixth row of Figure \ref{hinode2d}. In the degraded simulations and in the observed (strong) network regions, the $L/V$ ratio
exhibits a halo structure surrounding the strong flux concentrations in the
IGLs. At the spatial resolution of the coarse SP map (0\farcs64), the NCP (top row of Figure \ref{hinode2d}) does not show the same variation across individual features as in the simulations at full resolution or in high-resolution SP maps \citep{rezaei+etal2007}, but only exhibits positive NCP values roughly centered on the strongest circular polarization signals. Central areas with negative NCP apparently vanish at this spatial resolution.

Figure \ref{sp_polstat} displays histograms of several characteristic
polarization quantities for the two 630\,nm lines in the SP observations and the simulations degraded to SP resolution. The maximum polarization amplitude (top row, left) was used by \citet{khomenko+etal2005} to determine the value for the average magnetic flux density that best matches simulations and observations of the IN. The trend in the histograms with decreasing magnetic flux density is clear: the range of values in the histograms is compressed and the distributions shift to lower polarization degrees. The maximum of the distribution of the SP observation lies between the two simulation runs with low average magnetic flux, namely 20 G h and 50 G v. For the 100 G v and 200 G v simulation runs, the shape of the histograms of the polarization degree differs strongly from the observed one. Differences between the two 630\,nm lines (top row in Figure \ref{sp_polstat}) are minor. Whereas the polarization degree of most pixels clearly remains above the 3-$\sigma$ significance level, this is not the case for the linear polarization signal. 

\begin{figure*}
\resizebox{17.6cm}{!}{\includegraphics{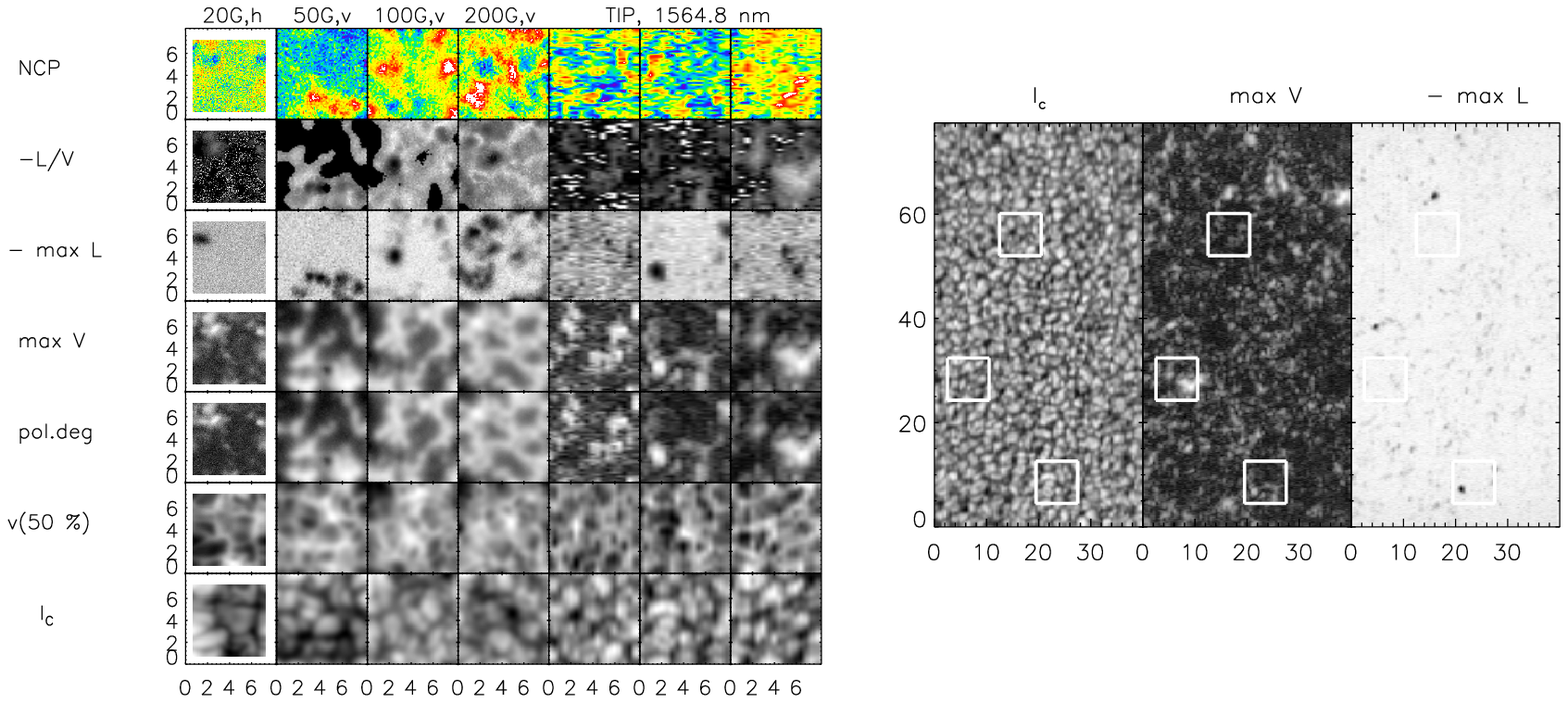}}
\caption{Left-hand panel: maps of polarization and intensity parameters
  for the 1564.8\,nm line in the TIP observations and the simulations degraded to TIP resolution. Same layout as in Figure \ref{hinode2d}. {Four leftmost 
    columns}: results based on degraded spectra of the 20 G h, 50 G v, 100 G v, and 200 G v runs. Fifth to seventh column: subfields in the TIP data corresponding to IN, network, and strong network. Right-hand panel: maps of continuum intensity (left subpanel), maximal $V$ amplitude (middle subpanel), and maximal linear polarization $L$ in inverted grey scale (right subpanel) for the 1564.8\,nm line in the full FOV of the TIP observations. The white squares indicate the subfields used in the left-hand panel. All tick marks are in arcsec.} \label{tip2d}
\end{figure*}

The dotted vertical line in the third and fourth panel of the top row of Figure \ref{sp_polstat} indicates the 3-$\sigma$ significance threshold in Stokes $Q$ for the SP data (see Table \ref{tab1}). Up to 75\,\% of the observed $Q$ signals are below this level and thus presumably correspond to noise \citep[see also][]{borrero+kobel2012}. The observed distributions of maximal Stokes $Q$ at 630.15\,nm and 630.25\,nm match those of the 20\,G h run for small polarization amplitudes below the 3-$\sigma$ level. For 630.25\,nm, the observed distribution is slightly above the 20\,G h run for polarization amplitudes larger than 3\,$\sigma$. All MHD runs with 50\,G or more are barely distinguishable from each other in the linear display. Their distributions are displaced towards higher polarization amplitudes relative to the observed distribution. Note that all MHD runs with 50\,G or more were performed with \textit{Stagger}, while the 20\,G h run was done with {CO$^5$BOLD}, so they were performed with different simulation codes and the corresponding initial setup.

The leftmost panel in the bottom row of Figure~\ref{sp_polstat} shows the distributions of maximal Stokes $Q$ amplitude for the line at 630.25\,nm with a logarithmic scale on the $y$-axis to increase the visibility of the differences. As already seen before, the observed distribution matches the distribution of the 20\,G h run at small polarization amplitudes, but stays above it for larger signals. The observed distribution stays below the distributions of all MHD runs with 50\,G or more between the 3-$\sigma$ level and an amplitude of about an 1\,\% (log$_{10}$ = -2) where it intersects the distributions of the 50 and 100\,G v runs. The distribution of the 200\,G v run stays above the observed distribution for all polarization amplitudes above the 3-$\sigma$ level.

In both the histograms of the zero-crossing velocity and the NCP (bottom row in Figure \ref{sp_polstat}) the observed distributions are broader than those of all simulation runs, with the 20 G h run as closest match. The average zero-crossing velocities after the spatial degradation are red-shifted in all cases \citep[cf.][]{grossmanndoerth+etal1996,sigwarth+etal1999,socasnavarro+etal2004,beck+etal2007}, but the observed values are displaced towards larger red-shifts than those of all simulations. In the ratio of $L/V$ (bottom rightmost panel in Figure \ref{sp_polstat}), the various simulation runs differ more clearly than in other parameters. Only the degraded 20\,G run matches the shape of the observed distribution of $L/V$, with a high fraction of pixels without any linear polarization signal (peak at zero) and a broad distribution with a local maximum around $L/V = 0.5$. The 100 G v and 200 G v runs fail to match the observed distribution (see also the sixth row from the bottom of Figure \ref{hinode2d}, considering that 90\,\% of the FOV correspond to the quiet IN subfield and not the two network cases). 

\subsection{TIP Observations and Degraded MHD Simulations at 1565\,nm}
We only discuss the TIP observations at 1565\,nm in similar detail as
the SP observations because they exhibit significant linear polarization
signals but at a higher S/N ratio. The corresponding maps and histograms for the GFPI and POLIS data are shown in Appendix \ref{appa} for completeness. Maps of the full FOV of these two observations can be found in \citet{puschmann+beck2011} and BE13.
\begin{figure}
\resizebox{8.8cm}{!}{\includegraphics{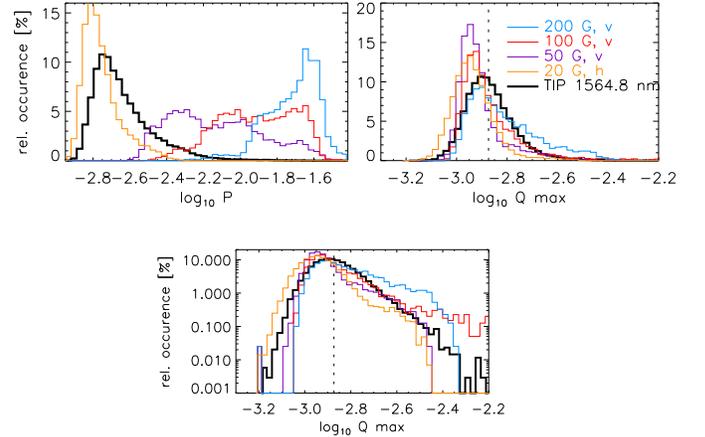}}
\caption{Histograms of maximum polarization degree (top left panel) and Stokes $Q$ amplitude (top right panel) for the 1564.8\,nm line. Bottom panel: histogram of Stokes $Q$ amplitude with logarithmic $y$-axis. Black lines: observations. Orange/purple/red/blue lines:  results based on degraded spectra of the 20 G h, 50 G v, 100 G v, and 200 G v runs.}\label{tip_polstat}
\end{figure}
\begin{figure*}
\centerline{\resizebox{12.cm}{!}{\includegraphics{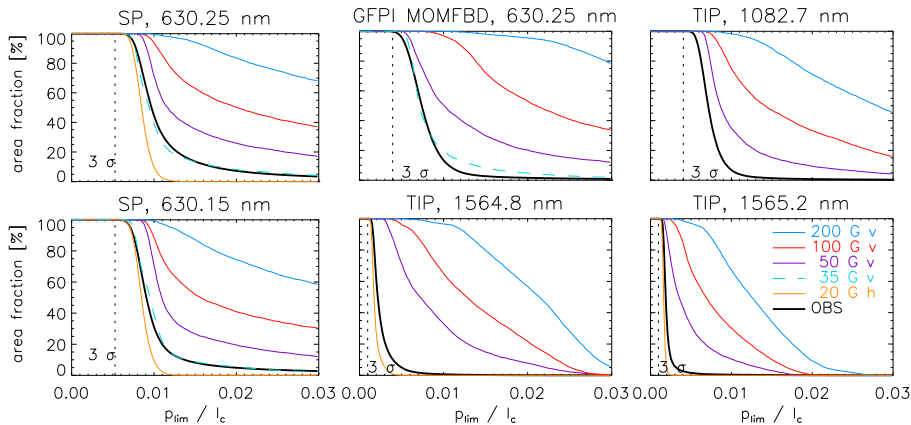}}}
\caption{Area fraction of profiles with a polarization degree above a given
  threshold. Instruments and spectral lines are indicated in the title of each panel. Black: observations. Orange/purple/red/blue lines: results based on degraded spectra of the 20 G h, 50 G v, 100 G v, and 200 G v runs. Turquoise dashed lines: artificial $\approx$35\,G run for the 630.15\,nm and 630.25\,nm lines.}\label{sp_cumupol}
\end{figure*}

The left panel of Figure \ref{tip2d} shows the degraded simulations and the corresponding subfields in the TIP observations at 1565\,nm. Most of the description for the SP data in the previous section also directly applies here. The most prominent difference is the level of linear polarization signals (fifth row from the bottom in Figure \ref{tip2d}). At the low noise level of the TIP data (see Table \ref{tab1}), several patches of linear polarization signal remain significant even in the degraded simulation runs. They are, however, again related to co-spatial vertical magnetic fields, and their relative area fraction exceeds the one in the quiet IN subfield of the observations. The area fraction of circular polarization signals in the IN subfield (fourth row in fifth column) is comparable to that of the degraded 20 G h and 50 G v runs (first and second column), but the latter shows some excess of polarization signals. 

In the histograms of the polarization degree (top left panel of
  Figure \ref{tip_polstat}) for the 1564.8\,nm line, the different simulation
  runs can clearly be distinguished, with a successive shift of the histograms
  to lower polarization degrees with decreasing magnetic flux density. The polarization amplitudes of the 50 G v run still significantly exceed those of the observations, whereas the distribution for the 20\,G h run is slightly off towards lower polarization degrees when compared with the observations. For the linear polarization signal, the picture is again not very clear in the linear display (top right panel of Figure \ref{tip_polstat}). The locations of the maxima of the distributions are nearly identical for all MHD runs in this case, also for the 20 G h run. The distributions only differ in the shape of the tail towards higher polarization amplitudes, which shows up more clearly in the plot that has a logarithmic scale for the $y$-axis (bottom panel of Figure \ref{tip_polstat}). The 100 and 200\,G v runs do not provide a good match to the observations at large polarization amplitudes. The distribution of the 50 G v run is intersected by the observed distribution at log$_{10} \approx -2.7$, while the distribution of the 20 G h run remains below the observed one for all amplitudes above the 3-$\sigma$ level that is indicated by the vertical dashed line.

\subsection{Cumulative Area Fraction above Polarization Threshold}
Figure \ref{sp_cumupol} shows the area fraction of profiles with a polarization degree above a given threshold for all but the POLIS data. The trend of polarization amplitudes with increasing magnetic flux density in the degraded simulation runs can clearly be distinguished, all simulation runs are well separated from each other (cf.~Figure \ref{stat2_fullres} for the simulations at full resolution). The 3-$\sigma$ significance level in the data as calculated from Table \ref{tab1} roughly coincides with the location where the area fraction starts to drop below 100\,\% because single noise peaks are likely to reach up to 3-$\sigma$ in each spectrum due to the several dozens wavelength pixels in each profile. For the SP data at 630.15\,nm and 630.25\,nm (left column of Figure \ref{sp_cumupol}) and the high-S/N TIP data (lower rightmost two panels), the observed curves lie between the 20 G h and 50 G v runs. For the other lines, all runs with 50 G or more show too large polarization amplitudes. 

We tried to achieve a better match between observations and simulations by removing part of the polarization signal from the spatially and spectrally degraded synthetic spectra of the simulation run that had 56\,G as average vertical magnetic flux density. We replaced the computed polarization degree in the lower 23 (out of 63) rows, where the largest polarization signals of the synthetic spectra are located (\textit{e.g.}, Figure \ref{tip2d}), by Gaussian noise with an appropriate amplitude. This removed about 40\,\% of the total area-integrated polarization signal $p_{\rm tot} = \int_A p\, dA$ from the FOV. Assuming that, in the WFL, magnetic flux density and induced polarization amplitude are directly proportional, the such-modified 50 G v run should then correspond to an average vertical magnetic flux density of about 35\,G across the full FOV. The resulting cumulative area fraction for these ``artificial'' spectra is superposed (dashed turquoise lines) in Figure \ref{sp_cumupol} on all 630\,nm data. The match to the SP observations (first column) is particularly good. It would be interesting to test this result by directly running a simulation with an average magnetic flux density of 35\,G, instead of relying, as we did here, on the downscaling of the total magnetic flux by removing part of the synthetic polarization signal from the FOV. Running a simulation with the new flux density value is, however, outside the scope of this work.

\section{Summary and Discussion}\label{sect_disc}
In a comparison of characteristic magnetic line parameters in observations and degraded MHD simulations of varying average magnetic flux density, we find that
\begin{itemize}
\item[a)] a good match to the observed spatial area coverage of polarization signals in either two-dimensional maps or the cumulative area fraction requires the magnetic field initially implanted in the simulations to have an average flux density between $20$\,G (horizontally-implanted magnetic field run) and $56$\,G (vertically-implanted magnetic field run),
\item[b)] the maxima of the histograms of the observed polarization degree $p$ and the curves of the cumulative area fraction lie between those of the simulation runs with a magnetic flux density of 20\,G (horizontally-implanted magnetic field run) and 56\,G (vertically-implanted magnetic field run),
\item[c)] the distributions of the observed Stokes $Q$ amplitudes lie above those of the 20\,G h run and primarily below those of the 100 and 200\,G v runs for amplitudes above the 3-$\sigma$ significance level,
\item[d)] the appearance of abnormal granulation in intensity maps of the degraded simulations with magnetic flux densities above 100\,G is to some extent only caused by a spatial smearing of BPs inside the IGLs with an otherwise rather regular granulation pattern in the simulations at full resolution,
\item[e)] linear polarization signals in the simulations are almost exclusively found in canopy-like structures surrounding strong vertical flux concentrations.
\end{itemize}

Using the characteristic properties of observed and synthetic spectra avoids ambiguities in the derivation of solar surface properties from observations. We find that an average magnetic flux density of about 35\,G matches synthetic spectra from degraded MHD simulations and observations in different spectral lines at a varying spatial resolution. The match is very good for the polarization degree that is dominated by the circular polarization, \textit{i.e.}, vertical magnetic fields, while the linear polarization signal can only be used to exclude the 20\,G h, 100 G v and 200\,G v runs as providing a good match. The value of 35\,G falls roughly in line with the 20\,G determined by \citet{khomenko+etal2005} in a similar comparison between simulations and observations, or with the list of observational results shown in Figure 3 of \citet{sanchezalmeida+martinezgonzalez2011}. An average magnetic flux density value in the simulation run of that order is, however, incompatible with values of close to or above 100\,G, as found by \citet{danilovic+etal2010} or \citet{orozcosuarez+katsukawa2012}. 

The spatial and spectral degradation of the simulations was taken into account in the determination of the best-matching simulation run. Any simulation run with an initial magnetic flux density significantly larger than $50$\,G, and hence also larger polarization amplitudes, would only provide a good match of simulations and observations for stronger spatial or spectral resolution effects than assumed here. Given the good match between the continuum intensity in observations and degraded simulations (Figures \ref{hinode2d}, \ref{tip2d}, \ref{gfpi_2d}, and \ref{polis_2d}) or between line parameters of the intensity spectra (BE13), stronger resolution effects seem not to be a valid option to reduce the polarization amplitudes in the simulations. 

The only way to encompass a significantly larger magnetic flux density would be to assume a strongly tangled magnetic field whose fingerprints would not be detectable at the relevant S/N and spatial resolution in linear \textit{and} circular polarization signals in the observed or degraded synthetic spectra. Note that also the temporal evolution in the atmosphere or simulation must not intensify this field to a detectable level. There are individual subfields inside the FOV that match the simulation runs with average magnetic flux densities of 100\,G or more, but their corresponding area coverage is (far) below 10\,\% of the observed full FOV (Figures \ref{sp_fullfov} and \ref{tip2d}).

At the noise level of our observations, most of the linear polarization signals in the simulations at full resolution disappear but for the high-S/N TIP data at 1565\,nm. The remaining linear polarization signals in the degraded simulations result from canopies of strong magnetic field concentrations as shown by the comparison of the degraded simulations with those at full spatial resolution or those at the spatial resolution of the MOMFBD GFPI spectra. The locations of large linear polarization amplitudes and large L/V ratios form a halo around the locations of maximal longitudinal magnetic flux with a reduction at the very center of the intergranular lanes. A similar, close relation between linear polarization signals (or likewise, horizontal magnetic fields) and circular polarization signals was found by \citet{ishikawa+tsuneta2011}. These kind of linear polarization signals are intrinsically different from the transient linear and circular polarization signals attributed to small-scale magnetic flux emergence events \citep[\textit{e.g.},][]{lites+etal1996,martinezgonzalez+etal2007,ishikawa+tsuneta2009,martinezgonzalez+bellotrubio2009,danilovic+etal2010a,gomory+etal2010,martinezgonzalez+etal2010,palacios+etal2012}. From the visual impression of the simulations at full resolution or after degradation, the simulations employed by us seem to lack such transient events, but still provide a reasonable reproduction of the spatial patterns of the observed linear polarization amplitudes.

Solar and stellar ``box-in-a-star'' 3D (magneto-)con-vection simulations have been able to achieve very satisfactory results in terms of matching a number of observational constraints. However, many more diagnostics remain to be tested, in particular with respect to the center-to-limb behavior of different spectral lines and to the interaction of radiation (including non-local effects), convection and magnetic fields. On top of that, detailed direct comparisons between physical and observable parameters obtained using an identical input setup for simulation runs performed with different solar magneto-convection codes are still lacking. Therefore, the relevant assumptions and approximations involved, \textit{e.g.}, diffusivities, radiative transfer and opacities, box size and resolution, or boundary conditions,  may cause that the results of the different codes have different levels of realism. \citet{beeck+etal2012} compared physical quantities and the center-to-limb variation of the continuum intensity obtained from \textit{field-free} solar convection simulations performed with different codes. However, for their comparison, they used snapshots obtained with different initial conditions, and they did not compare observable quantities nor any results (including Stokes parameters) based on MHD snapshots. Very recently, \citet[][A\&A, submitted]{CubasArmas+2017} performed a comparison using a more similar setup for two such codes, finding differences in terms of synthetic Stokes parameters that are likely due to the differing top boundary conditions.

Quiet Sun magnetism in IN regions is predicted to manifest itself at spatial scales and signal levels that are at the limit or below those reached by current spectropolarimetric instrumentation. The interpretation of spectropolarimetric observations of the Zeeman effect in the photosphere thus suffers from an interplay of magnetic field strength, total magnetic flux density, magnetic filling factor, stray light and noise. Average magnetic flux densities in the IN of above 100\,G were found by \citet{danilovic+etal2016} from the 2D inversion of synthetic observations and disk-center \textit{Hinode}/SP data in agreement with \citet{lites+etal2008}. The latter used a calibration curve from integrated observed polarization signal to magnetic flux density in the solar atmosphere that did not consider variations in the temperature stratification \citep[see][]{beck+rezaei2009}. In the case of \citet{danilovic+etal2016},  the magnetic field configuration in their MHD simulations is predominantly horizontal, with a transverse flux density of the order 50\,G that is 5--10 times larger than the longitudinal flux density.  \citet{danilovic+etal2010,danilovic+etal2016} multiplied the field strength by a factor of 2--3 in at least some of their MHD simulations to reach the observed level of IN spectropolarimetric signals. \citet{borrero+etal2011,borrero+kobel2012} studied the dependence of the inferred magnetic field inclination on the noise level of data and found that the presence of noise leads to an overestimation of the inclination. It is thus not clear at present if the IN really shows a dominant transversal/horizontal hG magnetic field or a longitudinal/vertical magnetic field of a few tens of G only \citep[\textit{e.g.},][]{beck+rezaei2009,ishikawa+tsuneta2011}. Our current results favor the latter since the observed circular and linear polarization signals from four different instruments can be matched with a predominantly vertical magnetic field of about 35\,G in the MHD simulations.

Spectropolarimetric observations at high spatial resolution \textit{and} with a high S/N \citep[\textit{e.g.,}][]{lites+etal2017} will be required to conclusively address the problem of quiet-Sun internetwork magnetic fields, with an improved evaluation of the data through the use of MHD simulations \citep[][]{rempel2014,riethmueller+solanki2017} or other analysis techniques \citep{vannoort2012,quinteronoda+etal2015}.

\section{Conclusions}\label{sect_concl}
An average vertical magnetic flux density of about 35\,G provides the best match of synthetic spectra from spatially and spectrally degraded MHD simulations and observed spectra in different spectral lines and at varying spatial resolution. MHD simulation runs with $50$\,G or more of unipolar vertical magnetic flux yield too strong polarization signals. The spatial patterns and relative area fraction of linear polarization signals in the simulations roughly match those in high-S/N observations, but in the simulations the linear polarization signals are clearly related to the canopy of magnetic field concentrations and not to transient magnetic flux emergence events. Any eventually existing strongly inclined or horizontal magnetic fields in the solar photosphere must be organized such as to avoid an increase in both circular or linear polarization amplitudes at the S/N level and spatial resolution of the observations used here. 

\begin{acknowledgements}
The VTT is operated by the Kiepenheuer-Institut f\"ur Sonnenphysik (KIS) at
the Spanish Observatorio del Teide of the Instituto de Astrof\'{\i}sica de
Canarias (IAC). The POLIS instrument has been a joint development of the High
Altitude Observatory (Boulder, USA) and the KIS. Hinode is a Japanese mission developed and launched by ISAS/JAXA, collaborating with NAOJ as a domestic partner, NASA and STFC (UK) as international partners. Scientific operation of the Hinode mission is conducted by the Hinode science team organized at ISAS/JAXA. This team mainly consists of scientists from institutes in the partner countries. Support for the post-launch operation is provided by JAXA and NAOJ (Japan), STFC (U.K.), NASA, ESA, and NSC (Norway). R.R. acknowledges support from the Spanish Ministry of Economy and Competitivity through project AYA2014-60476-P (Solar Magnetometry in the Era of Large Solar Telescopes). We would like to thank the referee for constructive suggestions.
\end{acknowledgements}
\bibliographystyle{aa}
\bibliography{references_luis_mod}
\begin{appendix}
\section{GFPI and POLIS data}\label{appa}
\begin{figure}
\centerline{\resizebox{17.6cm}{!}{\includegraphics{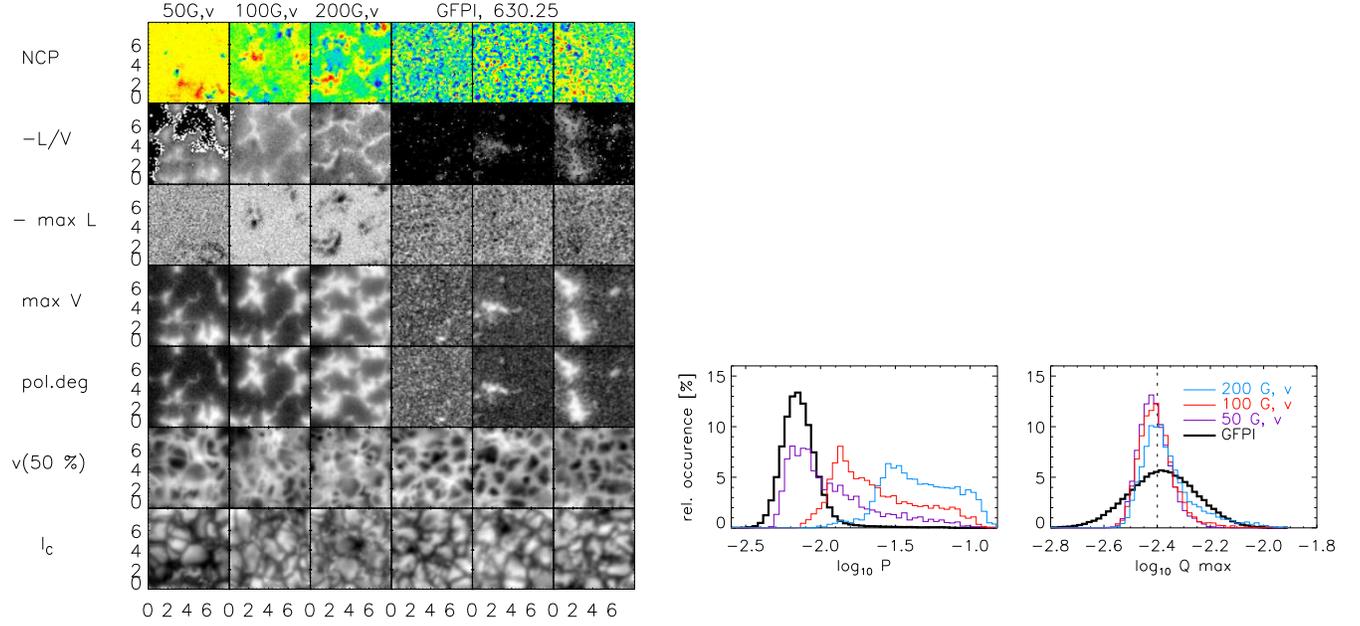}}}
\caption{Left-hand panel: maps of polarization and intensity parameters for the 630.25\,nm line in the GFPI observations and the simulations degraded to GFPI resolution. Same layout as in Figure \ref{hinode2d}. Tick marks are in arcsec. Right-hand panel: statistics of the polarization degree (left subpanel) and the maximal $Q$ amplitude (right subpanel) of the 630.25\,nm line. Black: observations. Purple/red/blue lines:  results based on degraded spectra of the 50 G v, 100 G v, and 200 G v runs. }\label{gfpi_2d}\label{gfpi_stat}
\end{figure}
For the comparison with the GFPI spectra, we did not degrade the synthetic spectra of the 20 G h run, because at the S/N of these observations the polarization signals would be below the detection limit. Even while for the GFPI the default noise rms is lower than for the SP data (see Table \ref{tab1}), we note that the rms value is misleading here, because the distribution of noise in the deconvolved spectra is non-Gaussian \citep{puschmann+beck2011}. With respect to POLIS, the SP data provide spectra with similar spectral characteristics at higher spatial resolution, implying that no additional information would be gained if we were to add a comparison of POLIS data and degraded spectra of the 20 G h run.

The left-hand panel of Figure \ref{gfpi_2d} shows the comparison between degraded spectra of the \textit{Stagger} MHD simulation runs and the GFPI data. At the spatial resolution of the GFPI, the linear polarization signals (fifth row from the bottom) in the simulations cleary trace the canopy of magnetic flux concentrations. The linear polarization signals above the central part of strong flux concentrations are nearly zero, but do form a halo with a radius of about 1$^{\prime\prime}$ around the locations of strongest circular polarization signal or vertical magnetic fields, respectively. In the statistics of the circular polarization signal (right-hand panel of Figure \ref{gfpi_stat}), the maximum of the observed distribution is displaced towards lower polarization amplitudes than for all simulation runs. The 50 G v run has the maximum of the distribution at about the correct location, but the shape of its distribution does not match the one of the observations because of its extended tail of high-amplitude polarization signals.
\begin{figure}
\centerline{\resizebox{17.6cm}{!}{\includegraphics{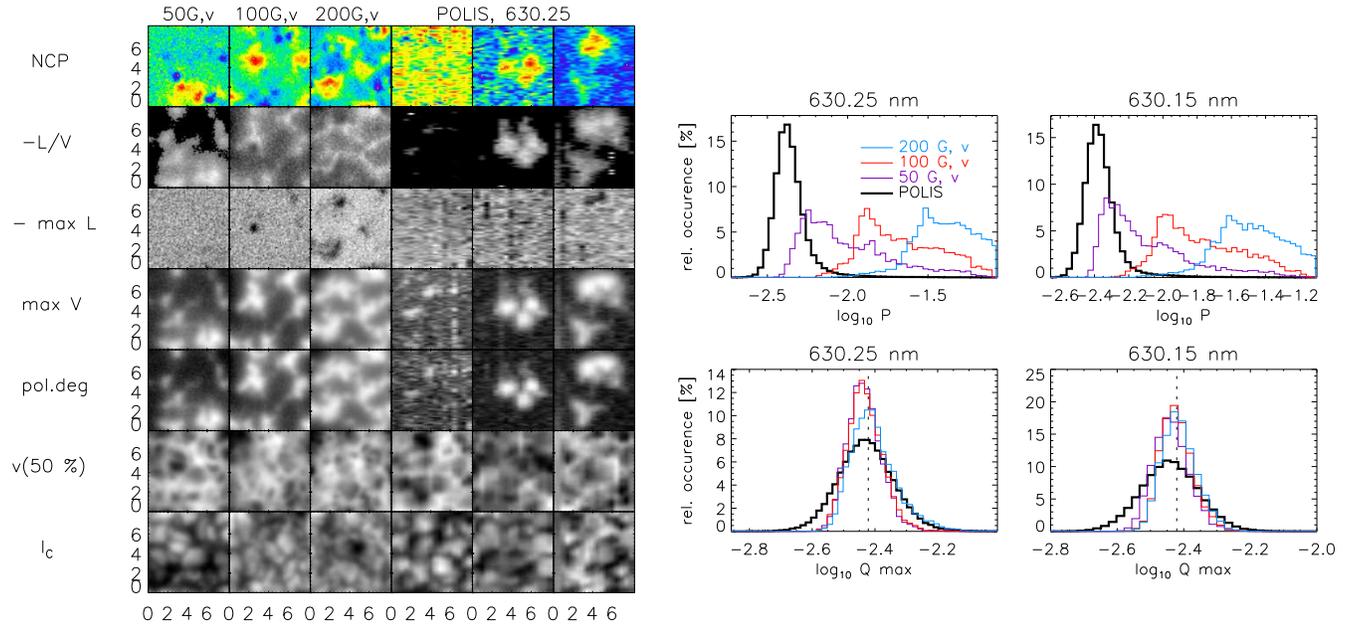}}}
\caption{Left-hand panel: maps of polarization and intensity parameters for the 630.25\,nm line in the POLIS observations and the simulations degraded to POLIS resolution. Same layout as in Figure \ref{hinode2d}. Tick marks are in arcsec. Right-hand panel: statistics of the polarization degree (top row) and the maximal Stokes $Q$ amplitude (bottom row) of the 630.25\,nm (left column) and 630.15\,nm (right column) lines. Black: observations. Purple/red/blue lines:  results based on degraded spectra of the 50 G v, 100 G v, and 200 G v runs.}\label{polis_2d}\label{polis_stat}
\end{figure}

The corresponding maps and statistics for the two 630\,nm lines observed with POLIS are shown in Figure \ref{polis_2d}. The distribution of the 50 G v run is slightly more displaced from the observed distribution than for the GFPI data (left subpanel in the right-hand panel of Figure \ref{gfpi_stat} and top left corner in the right-hand panel of Figure \ref{polis_stat}). Half of the observed linear polarization amplitudes in Stokes $Q$ (bottom row in the right-hand panel of Figure \ref{polis_stat}) are below the 3-$\sigma$ level. The different MHD runs can barely be distinguished from each other for small linear polarization amplitudes and again only differ in the tail towards large amplitudes. 
\end{appendix}
\end{document}